\documentstyle[12pt]{article}
\def \s{\sqrt{2}}

\begin{document}
\rightline{TECHNION-PH-94-7}
\rightline{EFI-94-14}
\rightline{UdeM-LPN-TH-94-192}
\rightline{March 1994}
\bigskip
\centerline{\bf WEAK COUPLING PHASE FROM DECAYS}
\centerline{\bf OF CHARGED B MESONS TO $\pi K$ AND
$\pi \pi$\footnote{Submitted to {\it Physical Review Letters}}}
\bigskip
\centerline{\it Michael Gronau}
\centerline{\it Department of Physics}
\centerline{\it Technion -- Israel Institute of Technology, Haifa 32000,
Israel}
\medskip
\centerline{and}
\medskip
\centerline{\it Jonathan L. Rosner}
\centerline{\it Enrico Fermi Institute and Department of Physics}
\centerline{\it University of Chicago, Chicago, IL 60637}
\medskip
\centerline{and}
\medskip
\centerline{\it David London}
\centerline{\it Laboratoire de Physique Nucl\'eaire}
\centerline{\it Universit\'e de Montr\'eal, Montr\'eal, PQ, Canada H3C 3J7}
\bigskip
\bigskip
\centerline{\bf ABSTRACT}
\medskip
\begin{quote}
The theory of $CP$ violation based on phases in weak couplings in the
Cabibbo-Kobayashi-Maskawa (CKM) matrix requires the phase $\gamma \equiv
{\rm Arg~} V^*_{ub}$ (in a standard convention) to be nonzero. A
measurement of $\gamma$ is proposed based on charged $B$ meson decay rates
to $\pi^+ K^0$, $\pi^0 K^+$, $\pi^+ \pi^0$, and the charge-conjugate
states. The corresponding branching ratios are expected to be of the order
of $10^{-5}$.
\end{quote}
\newpage

At present direct evidence for $CP$ violation comes exclusively from the
decays of neutral $K$ mesons. One theory of this phenomenon is based on
phases in the Cabibbo-Kobayashi-Maskawa (CKM) [1] matrix $V_{\alpha i}$,
which describes the weak charge-changing couplings of left-handed quarks $i
= (d,s,b)$ of charge $-1/3$ with left-handed quarks $\alpha = (u,c,t)$ of
charge $2/3$. By choosing five relative quark phases, one can take the
elements of $V$ along and just above the diagonal to be real (see, e.g.,
[2]). In this convention, taking account of the observed magnitudes of
elements, only $V_{ub}$ and $V_{td}$ can have significant nonzero phases.
The observed decays $K_L \to \pi^+ \pi^-$ and $\pi^0 \pi^0$ of the
long-lived neutral kaon and the charge asymmetry in semileptonic $K_L$
decays can be ascribed to a $CP$-violating mixing of $K^0$ and $\bar K^0$
arising from these phases. The CKM model of $CP$ violation also predicts
small differences in the ratios $\eta_{+-} \equiv A(K_L \to \pi^+ \pi^-)/
A(K_S \to \pi^+ \pi^-)$ and $\eta_{00} \equiv A(K_L \to \pi^0 \pi^0)/ A(K_S
\to \pi^0 \pi^0)$. Two recent experiments [3,4] reach different conclusions
about whether $\eta_{+-} = \eta_{00}$, and a satisfactory alternative
remains a ``superweak'' theory of direct $K^0 - \bar K^0$ mixing [5].

A fertile ground for testing the CKM model of $CP$ violation involves the
decays of mesons containing the fifth ($b$) quark [6]. Unequal rates for
decays of the mesons $B^0 \equiv \bar b d$ and $\bar B^0 \equiv b \bar d$
to $CP$ eigenstates like $J/\psi K_S$ can be interpreted crisply in terms
of the weak phase Arg $V_{td}$, without complications from strong
final-state interactions. However, the presence of $B^0 - \bar B^0$ mixing,
needed for the rate asymmetry, complicates the identification of neutral
$B$ mesons.

The decays of charged $B$ mesons can manifest $CP$ violation in the form of
unequal rates for such processes as $B^+ \to \pi^0 K^+$ and $B^- \to \pi^0
K^-$. While the charge of a $B$ meson is easily determined, strong
final-state interactions are required for such rate differences.
Differences in strong final-state phases among different eigenchannels are
expected to be small and uncertain. Thus, except in a few particular cases
[7], it has usually been assumed that information on CKM phases cannot be
extracted from the study of charged $B$ decays alone. Such decays can play
useful auxiliary roles in the separation of final-state interaction effects
from weak phases when decays of neutral $B$ mesons to $CP$ eigenstates are
also measured [8-10].

In this Letter we describe a way to obtain the weak phase $\gamma \equiv
{\rm Arg}~V^*_{ub}$ from the rates for the decays of charged $B$ mesons to
$\pi^+ K^0$, $\pi^0 K^+$, $\pi^+ \pi^0$, and the charge-conjugate states.
We expect equal rates for $B^+ \to \pi^+ \pi^0$ and $B^- \to \pi^- \pi^0$
on rather general grounds, and equal rates for $B^+ \to \pi^+ K^0$ and $B^-
\to \pi^- \bar K^0$ as a result of a specific assumption to be noted below.
The rates for $B^+ \to \pi^0 K^+$ and $B^- \to \pi^0 K^-$ can differ if
$CP$ is violated, but it is not necessary to measure a $CP$-violating
observable in order to obtain $\gamma$. The corresponding branching ratios
are expected to be of the order of $10^{-5}$, which is the level at which
decays of $B$ mesons to two light pseudoscalars have already been seen
[11].

The method relies upon an SU(3) relation between the amplitude for $B^+ \to
\pi^+ \pi^0$, which has isospin $I = 2$, and the isospin-3/2 amplitude in
$B \to \pi K$. Both amplitudes belong to the same 27-dimensional
representation of SU(3), and are related by a Clebsch-Gordan coefficient.
We perform the calculation using a convenient graphical method [12] which
has been shown equivalent to a general decomposition into SU(3)
representations. SU(3) breaking is also introduced, assuming that the
two-body hadronic decay amplitudes are factorizable. Other applications of
SU(3) to decays of $B$ mesons to pairs of light pseudoscalars have been
considered in Refs.~[13-15]. A more general recent discussion is contained
in Ref.~[16], where several new tests of the SU(3) assumption are
suggested.

The weak phase of the isospin-3/2 $\pi K$ amplitude is expected to be $\pm
\gamma$ for $B^{\pm}$ decays, while the strong phase does not change sign
under charge conjugation. The weak phases of the amplitude for $B^+ \to
\pi^+ K^0$ and $B^- \to \pi^- \bar K^0$ are both expected to be $\pi$ under
the assumption that weak annihilation graphs do not contribute to the
decay. (We shall suggest a test of this assumption.) Two triangle relations
satisfied by amplitudes, which include information from the rates for
$B^\pm \to \pi^0 K^\pm$, then allow one to separate out the desired weak
phase $\gamma$ modulo a discrete ambiguity.

We consider charmless decays of $B$ mesons to two light pseudoscalar mesons
within SU(3) [12,13]. Adopting the same conventions as Ref.~[12], we take
the $u,~d$, and $s$ quark to transform as a triplet of flavor SU(3), and
the $-\bar u,~\bar d$, and $\bar s$ to transform as an antitriplet. The
mesons are defined in such a way as to form isospin multiplets without
extra signs. Thus, the pions will belong to an isotriplet if we take
\begin{equation}
\pi^+ \equiv u \bar d~~,~~~\pi^0 \equiv (d \bar d - u \bar u)/\s~~,~~~
\pi^- \equiv - d \bar u~~~,
\end{equation}
while the kaons and antikaons will belong to isodoublets if we take
\begin{equation}
K^+ \equiv u \bar s~~,~~~K^0 \equiv d \bar s~~~,
\end{equation}
\begin{equation}
\bar K^0 \equiv s \bar d~~,~~~K^- \equiv - s \bar u~~~.
\end{equation}
The $B$ mesons are taken to be $B^+ \equiv \bar b u$, $B^0 \equiv \bar b
d$, and $B_s \equiv \bar b s$. Their charge-conjugates are defined as $B^-
\equiv - b \bar u$, $\bar B^0 \equiv b \bar d$, and $\bar B_s \equiv b \bar
s$.

The operators associated with the four-quark transition $\bar b \to \bar q
u \bar u$ and the direct (``penguin'') transition $\bar b \to \bar q$ ($q =
d$ or $s$) transform as a ${\bf 3^*,~6}$, or ${\bf 15^*}$ of SU(3). When
combined with the triplet of $B$ meson states, these operators lead to one
singlet, three octets and one ${\bf 27}$-plet, which appear in the
symmetric product of two octets (the pseudoscalar mesons, which are in an
S-wave final state). This leads to a decomposition of all
strangeness-preserving and strangeness-changing decay processes in terms of
five SU(3) reduced amplitudes.

As shown in Ref.~[12], this algebraic decomposition is equivalent to a
simpler graphical expansion. The six graphs which contribute are
illustrated in Fig.~1 [14]. They consist of a ``tree'' amplitude $T$
($T'$), a ``color-suppressed'' amplitude $C$ ($C'$), a ``penguin''
amplitude $P$ ($P'$), an ``exchange'' amplitude $E$ ($E'$), an
``annihilation'' amplitude $A$ ($A'$) and a ``penguin annihilation''
amplitude $PA$ ($PA'$). The unprimed amplitudes stand for
strangeness-preserving decays, while the primed ones represent
strangeness-changing processes. These amplitudes are related by simple CKM
factors. In particular:
\begin{equation}
T'/T = C'/C = E'/E = A'/A = r_u~~~,
\end{equation}
where $r_u\equiv V_{us}/V_{ud} \approx 0.23$. The set of six graphs is
overcomplete. They appear in all processes of the type $B\to PP$ in the
form of five linear combinations, corresponding to the five SU(3) reduced
matrix elements.

To apply SU(3) to the three decay processes, $B^+\to \pi^+\pi^0,~\pi^+
K^0,~ \pi^0 K^+$, we write the corresponding amplitudes in terms of their
graphical contributions:
\begin{equation}
\label{pipi}
A(B^+\to \pi^+\pi^0)=-{1\over\sqrt{2}}(T+C)~~~,
\end{equation}
\begin{equation}
\label{pikz}
A(B^+\to \pi^+ K^0)=P'+A'~~~,
\end{equation}
\begin{equation}
\label{pikp}
A(B^+\to \pi^0 K^+)=-{1\over\sqrt{2}}(T'+C'+P'+A')~~~.
\end{equation}
Here, for instance, the combinations $C'+T'$ and $P'+A'$ form two of the
five linearly independent combinations of graphical contributions. We
immediately find:
\begin{equation}
\label{tri}
\sqrt{2} A(B^+ \to \pi^0 K^+) + A(B^+ \to \pi^+ K^0) = r_u \sqrt{2}
A(B^+\to \pi^+\pi^0)~~~.
\end{equation}
This relation is described by a triangle in the complex plane, as shown in
Fig.~2. The corresponding triangle for the charge-conjugate process is also
shown. Notice that the two triangles share one side. An SU(3) assumption
has been made in order to obtain this simple result.

The diagrams denoted by $E,~A,~PA$ involve contributions to amplitudes
which should behave as $f_B/m_B$ in comparison with those from the diagrams
$T,~C$, and $P$ (and similarly for their primed counterparts).  This
suppression is due to the smallness of the $B$ meson wave function at the
origin, and it should remain valid unless rescattering effects are
important. Such rescatterings indeed could be responsible for certain
decays of charmed particles (such as $D^0 \to \bar K^0 \phi$), but should
be less important for the higher-energy $B$ decays. In addition the
diagrams $E$ and $A$ are also helicity suppressed by a factor
$m_{u,d,s}/m_B$ since the $B$ mesons are pseudoscalars.

If rescattering effects are small and the amplitudes $A,~E$, and $PA$ can
be neglected, the rate for $B^0\to K^+ K^-$ will be suppressed relative to
$B^0\to\pi^+\pi^-$, since the amplitudes for these processes are given by
\begin{equation}
A(B^0\to\pi^+\pi^-)=-(T+P+E+PA)~~~,
\end{equation}
\begin{equation}
A(B^0\to K^+ K^-)=-(E+PA)~~~.
\end{equation}

Assuming that the amplitude $A'$ can be neglected in (\ref{pikz}) and
(\ref{pikp}), the phases in the decay amplitudes and those for the
charge-conjugate processes have simple relations to one another. The phase
of the $P'$ amplitude, which is expected to be dominated by the top quark
loop [17], should be approximately Arg $V^*_{tb}V_{ts} = \pi$. Then we may
denote
\begin{equation}
A(B^+ \to \pi^+ K^0) = A(B^- \to \pi^- \bar K^0) = P' = - a_P
e^{i \delta_P}~~~,
\end{equation}
where $a_P$ is real. Note that the rates for the process and its charge
conjugate are equal. Similarly, taking account of the factor which relates
$T + C$ to $T' + C'$ and using Arg $V^*_{ub}V_{us} = \gamma$, we find
\begin{equation}
r_u \sqrt{2} A(B^+ \to \pi^+ \pi^0) = -(T'+C') = a_T e^{i \delta_T}
e^{i \gamma}~~~,
\end{equation}
while
\begin{equation}
r_u \sqrt{2} A(B^- \to \pi^- \pi^0) = a_T e^{i \delta_T} e^{-i \gamma}~~~,
\end{equation}
with $a_T$ real. The rates for these two processes are equal because they
involve a single weak phase and a single strong phase. The difference in
phase between these two amplitudes is just $2 \gamma$.

The third side of each amplitude triangle is provided by the rate for the
decay $B^+ \to \pi^0 K^+$ or $B^- \to \pi^0 K^-$, as shown in Fig.~2. Here
$A^{0+} \equiv A(B^+ \to \pi^0 K^+)$, $A^{+0} \equiv A(B^+ \to \pi^+ K^0)$,
$A^{0-} \equiv A(B^- \to \pi^0K^-)$, $A^{-0} \equiv A(B^- \to \pi^- \bar
K^0)$, $A_{\pi \pi}^{+0} \equiv A(B^+ \to \pi^+ \pi^0)$, $A_{\pi \pi}^{-0}
\equiv A(B^- \to \pi^- \pi^0)$. Modulo a two-fold ambiguity which
corresponds to flipping one triangle about the horizontal axis, the rates
determine the shapes of the triangles and hence the difference $2 \gamma$.
The flipping of one triangle corresponds to interchanging $\gamma$ and
$\delta_P-\delta_T$. In general, $CP$ violation is expected to show up as a
difference in rates between $B^+ \to \pi^0 K^+$ and its charge-conjugate,
since two CKM amplitudes with different phases interfere in this process.
The crucial point in determining $\gamma$ is that the magnitudes of these
two amplitudes are separately measured in $B^+\to\pi^+ K^0$ and
$B^+\to\pi^+\pi^0$. If $\delta_P - \delta_T = 0$, we will not observe such
a difference in rates. In that case, however, we would have to choose the
lower of Figs.~2, since only this configuration would correspond to a
nonzero value of $\gamma$.

One can take account of SU(3) breaking in factorizable amplitudes by noting
that the decay $B^+ \to \pi^+ \pi^0$ involves a factor of the pion decay
constant $f_\pi$, whereas the $I = 3/2$ amplitude in $B \to \pi K$ should
involve a factor $f_K$. Thus, one should probably multiply $r_u$ in all the
relations presented here by the factor $f_K/f_\pi \approx 1.2$. This
prescription was adopted in Ref.~[15].

Fig.~2 will permit the measurement of $\gamma$ if each of the decay rates
can be measured with sufficient accuracy. Explicitly, defining $a \equiv
|A^{+0}| = |A^{-0}|$, $b \equiv (f_K/f_{\pi}) r_u \sqrt{2}|A^{+0}_{\pi
\pi}| = (f_K/f_{\pi}) r_u \sqrt{2}|A^{-0}_{\pi \pi}|$, $c \equiv
\sqrt{2}|A^{0+}|$, $c' \equiv \sqrt{2}|A^{0-}|$, one has
\begin{equation}
4ab \sin \gamma = \pm \{[(a+b)^2 - c^2][c'^2 - (a-b)^2]\}^{1/2}
\pm \{c \leftrightarrow c'\}~~~.
\end{equation}

The present data on $B^0$ decays to pairs of pseudoscalars [11] do not
allow one to distinguish between $\pi^- K^+$ and $\pi^+ \pi^-$ final
states. The combined branching ratio is about $2 \times 10^{-5}$, with
equal rates for $\pi^- K^+$ and $\pi^+ \pi^-$ being most likely. If this is
true, the amplitudes $T$ and $P'$ have about the same magnitude, so that
the short sides of the triangles in Fig.~2 are probably about 1/4 to 1/3
($\approx (f_K/f_{\pi})r_u$) the lengths of the other two sides. Then the
``long'' sides of the triangle must be measured with fractional accuracies
of about $(f_K/f_{\pi})r_u \delta \gamma$ in order to achieve an accuracy
of $\delta \gamma$ in the angle $\gamma$. For example, to measure $\gamma$
to a statistical accuracy of about $10^{\circ}$, one probably needs
fractional errors of about 1/20 in amplitudes, or 10\% in rates. This would
require at least 100 decays in each channel of interest.

We end with some comments about other ways of measuring weak phases.

(1) Another measurement of $\gamma$ from charged $B$ decays uses the
processes $B^{\pm}\to K^{\pm}D^0,~\to K^{\pm} \overline{D}^0,~ \to
K^{\pm}D_{CP}$, where $D_{CP}$ denotes a $CP$ eigenstate [7,18]. The three
$B^+$ amplitudes and their charge-conjugates obey two triangle relations
similar to the above. Here too the angle $\gamma$ can be measured without
an observation of $CP$ violation in $B^{\pm}\to K^{\pm} D_{CP}$, even when
the final-state phase differences are too small to detect.  While $B^+\to
K^+ D^0$ may be strongly color-suppressed, all the measured rates are
expected to be of comparable magnitudes in the method presented here.

(2) The present method uses $B$ decay modes with rates similar to $B^0\to
\pi^+ \pi^-$ decays. The use of $\pi^+\pi^-$ decays requires tagging the
neutral $B$ meson flavor at time of production, and suffers from
uncertainties associated with penguin amplitudes [8]. These uncertainties
can be eliminated by a complete isospin analysis of all charge states in
$B\to \pi\pi$ decays [9], or at least estimated by relating via SU(3) the
rates of $B^0 \to \pi^+\pi^-$ and $B^0\to\pi^- K^+$ [15]. Information from
additional $\pi \pi,~\pi K$, and $K \bar K$ branching ratios of charged and
neutral $B$'s can be combined with the rates mentioned here to further
eliminate ambiguities and constrain other weak phases [16].

To summarize, we have shown that measurements of the rates for charged $B$
decays to $\pi K$ and $\pi \pi$, together with a simple SU(3) relation,
suffice to specify the geometry of amplitude triangles from which one can
extract the weak phase $\gamma = {\rm Arg}~V^*_{ub}$, where $V_{ub}$
describes an element of the Cabibbo-Kobayashi-Maskawa (CKM) matrix. No
final-state-interaction phases need be specified. A non-zero value of
$\gamma$ in accord with other analyses of parameters in the CKM matrix
would provide valuable confirmation of a popular model of $CP$ violation.
\bigskip

\centerline{\bf ACKNOWLEDGMENTS}
\bigskip

We thank B. Blok for fruitful discussions. M. Gronau and J. Rosner
respectively wish to acknowledge the hospitality of the Universit\'e de
Montr\'eal and the Technion during parts of this investigation. This work
was supported in part by the United States -- Israel Binational Science
Foundation under Research Grant Agreement 90-00483/2, by the German-Israeli
Foundation for Scientific Research and Development, by the Fund for
Promotion of Research at the Technion, by the United States Department of
Energy under Contract No. DE FG02 90ER40560, and by the N. S. E. R. C. of
Canada and les Fonds F. C. A. R. du Qu\'ebec.
\newpage

\centerline{\bf REFERENCES}
\begin{enumerate}

\item[{[1]}] N. Cabibbo, Phys.~Rev.~Lett.~{\bf 10}, 531 (1963); M.
Kobayashi and T. Maskawa, Prog.~Theor.~Phys.{\bf 49}, 652 (1973).

\item[{[2]}] J. D. Bjorken and I. Dunietz, Phys.~Rev.~D {\bf 36}, 2109
(1987).

\item[{[3]}] Fermilab E731 Collaboration, L. K. Gibbons {\it et al.},
Phys.~Rev.~Lett~{\bf 70}, 1203 (1993).

\item[{[4]}] CERN NA31 Collaboration, G. D. Barr {\it et al.,}
Phys.~Lett.~B {\bf 317}, 233 (1993).

\item[{[5]}] L. Wolfenstein, Phys.~Rev.~Lett.~{\bf 13}, 562 (1964).

\item[{[6]}] See, e.g., Y. Nir and H. Quinn, Ann.~Rev.~Nucl.~Part.~Sci.~
{\bf 42}, 211 (1992).

\item[{[7]}] M. Gronau and D. Wyler, Phys.~Lett.~B {\bf 265}, 172 (1991).
See also M. Gronau and D. London, Phys.~Lett.~B {\bf 253}, 483 (1991); I.
Dunietz, Phys.~Lett.~B {\bf 270}, 75 (1991).

\item[{[8]}] D. London and R. Peccei, Phys.~Lett.~B {\bf 223}, 257 (1989);
M. Gronau, Phys.~Rev.~Lett.~{\bf 63}, 1451 (1989); B. Grinstein,
Phys.~Lett.~B {\bf 229}, 280 (1989); M. Gronau, Phys.~Lett.~B {\bf 300},
163 (1993).

\item[{[9]}] M. Gronau and D. London, Phys.~Rev.~Lett.~{\bf 65}, 3381
(1990).

\item[{[10]}] Yosef Nir and Helen R. Quinn, Phys.~Rev.~Lett.~{\bf 67}, 541
(1991); Michael Gronau, Phys.~Lett.~B {\bf 265}, 389 (1991); H. J. Lipkin,
Y. Nir, H. R. Quinn and A. E. Snyder, Phys.~Rev.~D {\bf 44}, 1454 (1991).

\item[{[11]}] M. Battle {\it et al.} (CLEO Collaboration),
Phys.~Rev.~Lett.~{\bf 71}, 3922 (1993).

\item[{[12]}] D. Zeppenfeld, Z. Phys.~C {\bf 8}, 77 (1981).

\item[{[13]}] M. Savage and M. Wise, Phys.~Rev.~D {\bf 39}, 3346 (1989);
{\it ibid.}~{\bf 40}, 3127(E) (1989).

\item[{[14]}] L. L. Chau {\it et al.}, Phys.~Rev.~D {\bf 43}, 2176
(1991).

\item[{[15]}] J. Silva and L. Wolfenstein, Phys.~Rev. D {\bf 49}, R1151
(1994).

\item[{[16]}] M. Gronau, O. Hernandez, D. London, and J. L. Rosner,
Technion preprint TECHNION-PH-94-8, March, 1994, to be submitted to
Phys.~Rev.~D.

\item[{[17]}] T. Inami and C. S. Lim, Prog.~Theor.~Phys.~{\bf 65}, 297,
1772(E) (1981); G. Eilam and N. G. Deshpande, Phys.~Rev.~D {\bf 26}, 2463
(1982).

\item[{[18]}] S. L. Stone, in {\it Beauty 93}, Proceedings of the First
International Workshop on B Physics at Hadron Machines, Liblice Castle,
Melnik, Czech Republic, January 1993, ed. P. E. Schlein, Nucl. Instrum.
Meth. {\bf 33}, 15 (1993).

\end{enumerate}
\bigskip

\centerline{\bf FIGURE CAPTIONS}
\bigskip

\noindent
FIG.~1. Diagrams describing decays of $B$ mesons to pairs of light
pseudoscalar mesons. Here $\bar q = \bar d$ for unprimed amplitudes and
$\bar s$ for primed amplitudes.  (a) ``Tree'' (color-favored) amplitude $T$
or $T'$; (b) ``Color-suppressed'' amplitude $C$ or $C'$; (c) ``Penguin''
amplitude $P$ or $P'$ (we do not show intermediate quarks and gluons); (d)
``Exchange'' amplitude $E$ or $E'$; (e) ``Annihilation'' amplitude $A$ or
$A'$; (f) ``Penguin annihilation'' amplitude $PA$ or $PA'$.
\bigskip

\noindent
FIG.~2. SU(3) triangles involving decays of charged $B$'s which may be used
to measure the angle $\gamma$. Here $A^{0+} \equiv A(B^+ \to \pi^0 K^+)$,
$A^{+0} \equiv A(B^+ \to \pi^+ K^0)$, $A^{0-} \equiv A(B^- \to \pi^0K^-)$,
$A^{-0} \equiv A(B^- \to \pi^- \bar K^0)$, $A_{\pi \pi}^{+0} \equiv A(B^+
\to \pi^+ \pi^0)$, $A_{\pi \pi}^{-0} \equiv A(B^- \to \pi^- \pi^0)$. The
lower figure shows one of the triangles flipped about the horizontal axis.
This solution must be chosen when $|A^{0+}| = |A^{0-}|$ if $\gamma \ne 0$.

\end{document}